\documentclass[12pt]{article}
\usepackage{graphicx}

\def\institute{$^1$Centre for Detector and Related Software Technology,\\
Department of Physics and Astrophysics, University of Delhi, INDIA}
\def\support{\footnote{Work supported by the Department of Atomic Energy and the Department of Science and Technology, India, under collaboration with CMS-CERN.}}
\def\Title#1{\begin{center} {\Large #1 } \end{center}}
\def\Author#1{\begin{center}{ \sc #1} \end{center}}
\def\Address#1{\begin{center}{ \it #1} \end{center}}

\newenvironment{Abstract}{\begin{quotation}  }{\end{quotation}}
\newenvironment{Presented}{\begin{quotation} \begin{center} 
             PRESENTED AT\end{center}\bigskip 
      \begin{center}\begin{large}}{\end{large}\end{center} \end{quotation}}

\def\beq{\begin{equation}}
\def\eeq#1{\label{#1}\end{equation}}
\def\eeqn{\end{equation}}

\def\beqa{\begin{eqnarray}}
\def\eeqa#1{\label{#1}\end{eqnarray}}
\def\eeqan{\end{eqnarray}}

\let\bar=\overbar

\def\Dslash{\not{\hbox{\kern-4pt $D$}}}
\def\dslash{\not{\hbox{\kern-2pt $\del$}}}

\def\msb{{\bar{\ssstyle M \kern -1pt S}}}

\begin{document}
\begin{titlepage}

\vfill
\Title{First measurement of tW production cross-section at $\sqrt s$~= 13 TeV with CMS}
\vfill
\Author{ Priyanka$^*$\support, Kirti Ranjan$^1$, Ashutosh Bhardwaj$^1$\\on behalf of CMS collaboration}
\Address{\institute}
\vfill
\begin{Abstract}
The inclusive cross-section for tW production in proton-proton collisions at $\sqrt{s} = 13$ TeV is measured for an integrated luminosity of 35.9 fb$^{-1}$ collected by the CMS experiment. The measurement is performed using events with one electron and one muon in the final state and at least one b-quark jet, and utilises kinematic differences between the signal and the dominating $t\bar{t}$ background using multivariant discriminants which is designed to disentangle the two processes. The measured cross-section of $\sigma = 63.1 \pm 1.8~({\rm stat}) \pm 6.4~({\rm syst}) \pm 2.1~({\rm lumi})$ pb is observed to be in agreement with the Standard Model.
\end{Abstract}
\vfill
\begin{Presented}
$11^\mathrm{th}$ International Workshop on Top Quark Physics\\
Bad Neuenahr, Germany, September 16--21, 2018
\end{Presented}
\vfill
\end{titlepage}
\def\thefootnote{\fnsymbol{footnote}}
\setcounter{footnote}{0}

\section{Introduction}
The top quark was first discovered by the CDF $\&$~the D0 collaborations in pair production mode, through strong interaction at the Tevatron in 1995 \cite{Ref1,Ref2}, which is the dominant production mode. Top quark has been the heaviest elementary particle found till date in the standard model (SM) of particle physics. Its short lifetime makes it unique as it decays into a W boson and a b quark before hadronizing. Apart from the strong production it is also produced via electroweak production, also known as single-top quark production. This was obeserved in 2009 by the CDF and D0 collaborations at the Tevatron \cite{Ref3,Ref4}. Single-top quark allows probing of BSM and it also becomes a background to the precision $t\bar t$~physics and to many other SM processes. The production of single-top quark happens via three subprocesses: t-channel, the dominant one at the LHC; the s-channel, the least dominant at the LHC; and the associated production of top quark and W boson, also known as the tW-channel, which is the second most dominant process at the LHC.
\begin{figure}[h]
\centering
\includegraphics[width=7cm]{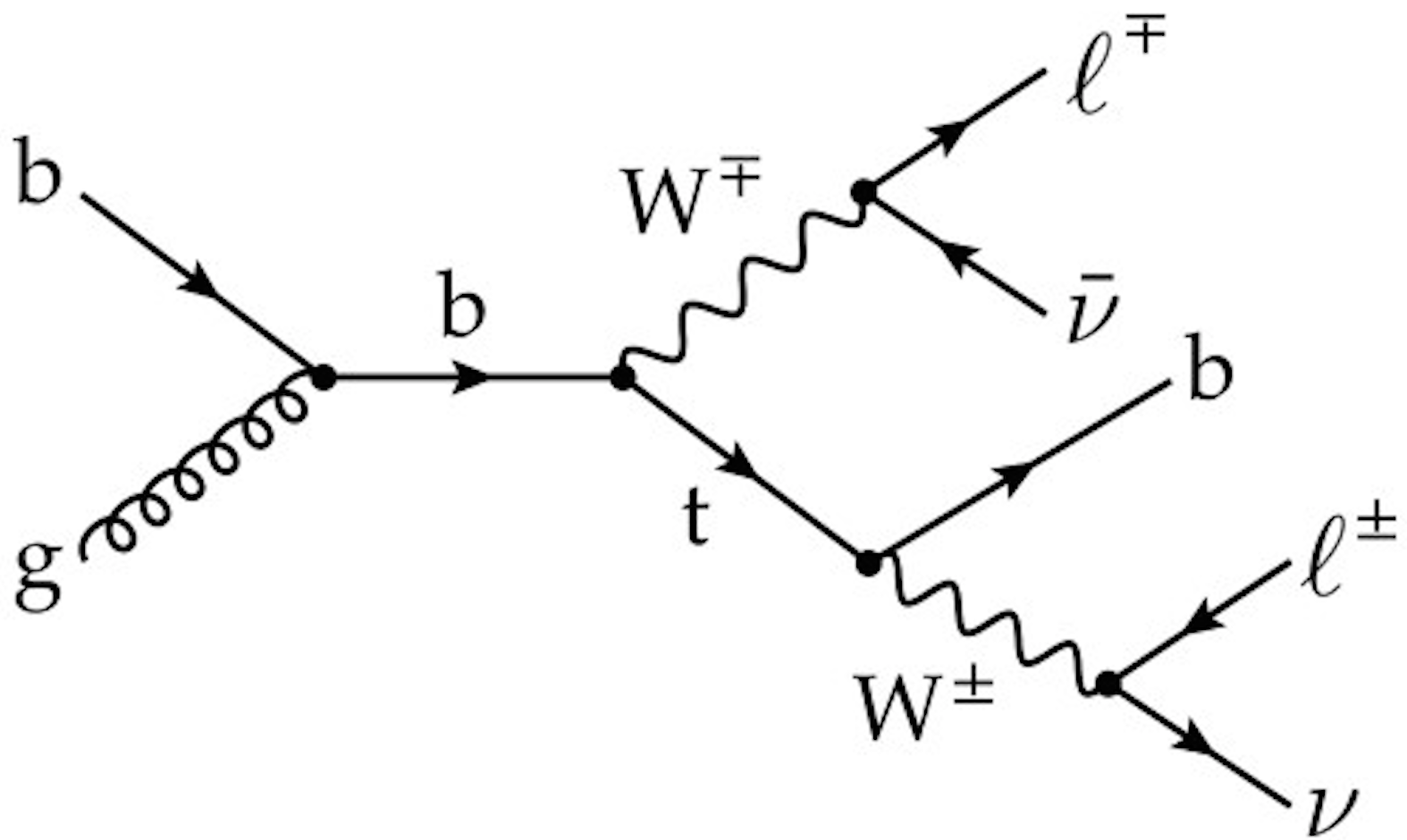}
\caption{Leading order Feynman diagram with signature event seletion.}
\label{fig-1}       
\end{figure}
However, there exists a major difficulty in the observation of tW subprocess as it interferes with the strong production of top quark pair production ($t\bar t$) at next-to-leading order (NLO). To deal with this difficulty at NLO, two methods are generally considered: diagram Removal (DR) and the diagram subtraction (DS), which removes the $t\bar t$~effect at the amplitude level and at the cross-section level respectively \cite{Ref5}. Evidence for the tW production was found at centre-of-mass energy ($\sqrt s =$) 7 TeV with $2.05 fb^{-1}$~of data at ATLAS experiment and its observation was found at $\sqrt s =$~8 TeV with $12.2 fb^{-1}$~of data at CMS experiment.
\section{Event Selection and Results}
\label{tW-channel}
\begin{figure}[h]
\centering
\includegraphics[width=7cm,clip]{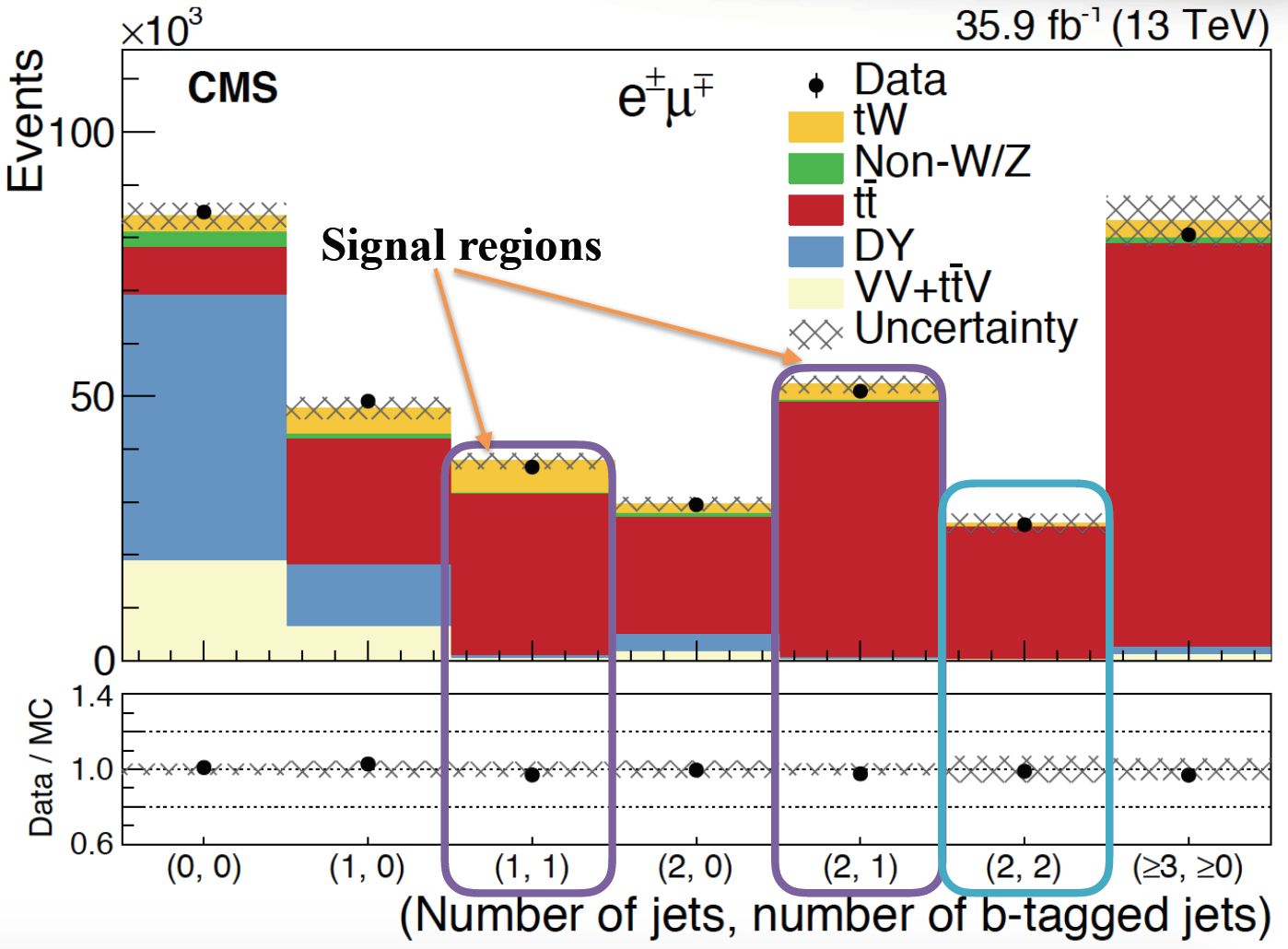}
\caption{Comparison of yields observed in the data with those expected from simulation as a function of the number of jets and the number of b-tagged jets, for the events satisfying the baseline dilepton selections. The uncertainity band here includes the statistical and systematic uncertainties except those arising from background normalization. The bottom panel shows the ratio of the data to the sum of the expected yields.}
\label{fig-2}       
\end{figure}
\begin{figure}[h]
\centering
\includegraphics[width=7cm,clip]{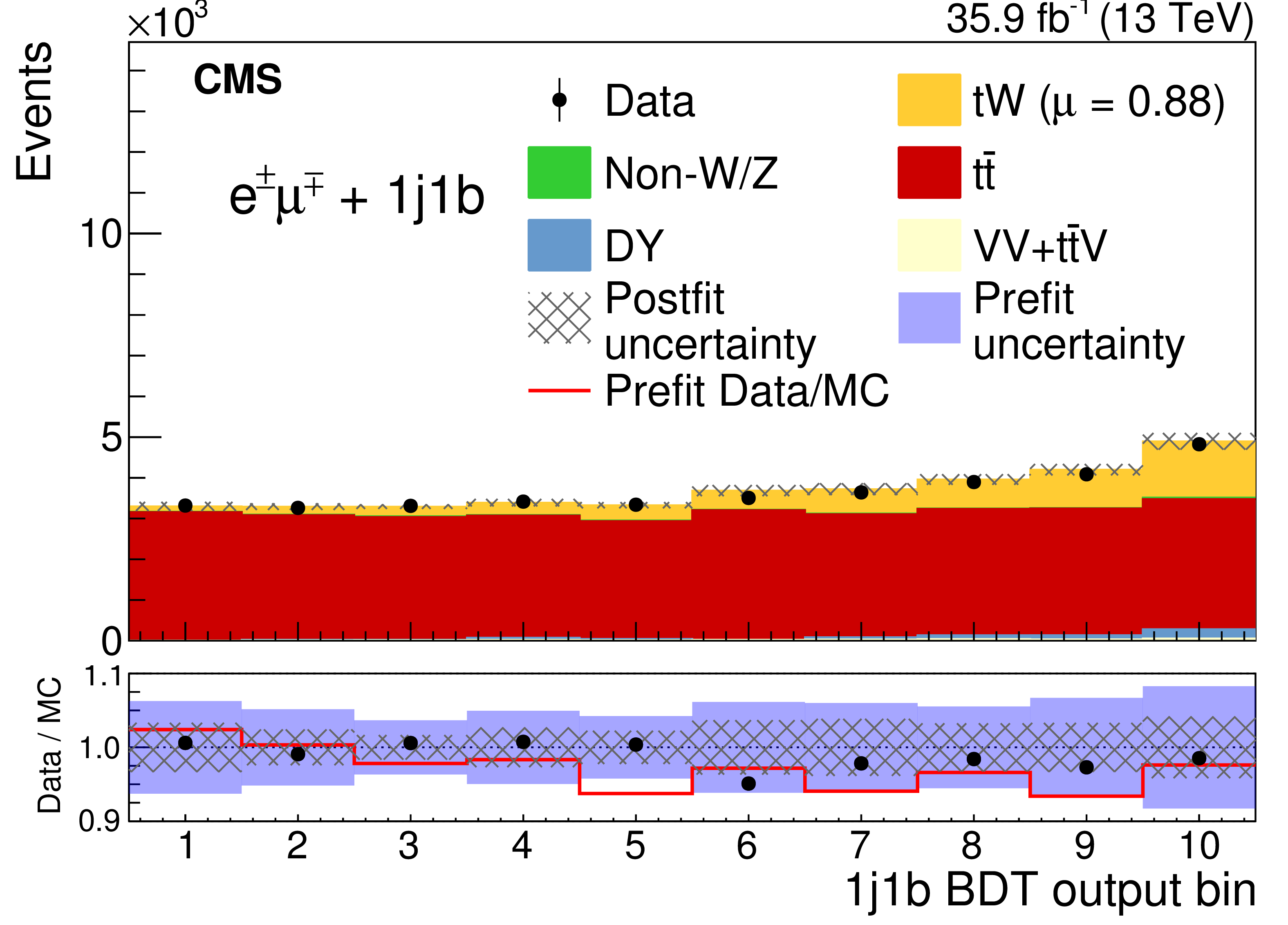}
\includegraphics[width=7cm,clip]{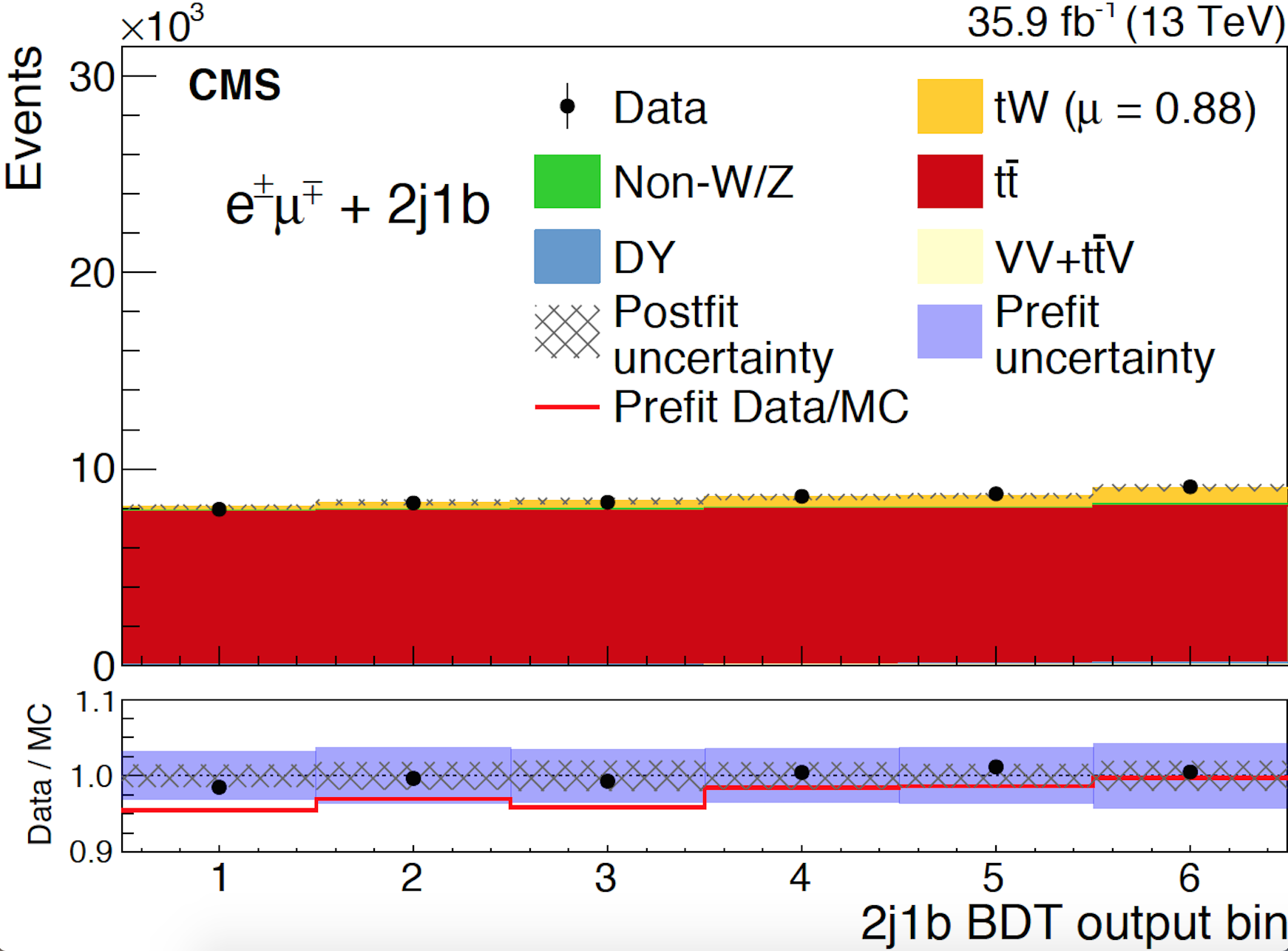}
\caption{BDT output in the 1jet 1b-tag region after the fit is performed for the observed data is compared with the simulated events (left). Comparison of the BDT output in the 2jets 1b-tag region after the fit is performed for the observed data and the simulated events (right). The uncertainity band in the plot includes the statistical and systematic uncertainties. The bottom panel shows the ratios of data to the expectation from simulations and also from the fit.}
\label{fig-3}       
\end{figure}
At CMS, the first measurement of the tW production cross-section was performed at $\sqrt s =$~of 13 TeV using the full 2016 data corresponding to an integrated luminosity of 35.9 $fb^{-1}$ \cite{Ref6}. The signature events of this analysis contains the events with 2 opposite sign leptons which can be electrons or muons, and 1 jet which originates from a bottom quark (Figure~\ref{fig-1}). However, we defined 3 regions for the signal extraction to increase the signal content which are 1 jet also tagged as b-jet (1j1t) and 2 jets with either 1 or both being tagged as b-jets (2j1t or 2j2t) as shown in the enclosed boxes in Figure~\ref{fig-2}. The main backgrounds to this channel come from $t\bar t$ which is the dominant one, $Z+jets$, $W+jets$, $ZZ$, $WW$, $WZ$~and $WW$. Evaluation of the backgrounds that may contaminate the signal was performed on Drell-Yan and non-W backgrounds using data-driven technique. Simultaneous maximum likelihood fit to the Boosted Decision Tree (BDT) distributions in 1j1t and 2j1t regions and sub-leading jet $p_T$~distribution in 2j2t region was performed for the signal extraction. Figure~\ref{fig-3} shows the 1j1t and 2j1t BDT distributions respectively. Systematic uncertainties are parametrized as nuisance parameters of the fit. Jet energy scaling, lepton identification and $t\bar t$~modeling contribute to the main systematic uncertainties. The cross section of the associated production of top quark with W boson is measured by the CMS as $\sigma =63.1\pm 1.8(stat.)\pm 6.4(syst.)\pm 2.1(lumi) pb$, achieving a relative uncertainty of 11$\%$, which is in agreement with the SM prediction of $\sigma (NNLO) = 71.7\pm 1.8(scale)\pm 3.4(PDF) pb$. At $\sqrt s =$~13 TeV this is the first measurement of the $tW$~process by the CMS Collaboration performed using BDT analysis. Measured cross-section is also in agreement with a similar measurement by ATLAS Collaboration.
\section{Summary}
CMS has performed the precision measurement of tW-channel using the p-p collision data at $\sqrt s =$~13 TeV. Increased luminosity has befitted the measurement and the results are in agreement with SM.

\end{document}